\documentstyle[12pt]{article}
\begin{document}
%
%\textwidth = 12truecm
%\textheight =  17truecm
%% Title and abstract page %%
\begin{titlepage}

\begin{center}
{\Large\bf The Hamiltonian Formulation}

\vspace{5pt}

{\Large\bf of}

\vspace{5pt}

{\Large\bf Higher Order
Dynamical Systems}

\end{center}

\vskip 25mm

\begin{center}
Jan Govaerts\\
{\em Institut de Physique Nucl\'eaire}\\
{\em Universit\'e Catholique de Louvain}\\
{\em B-1348 Louvain-la-Neuve, Belgium}\\
\vspace{15pt}
and\\
\vspace{15pt}
Maher S. Rashid\\
{\em International Center for Theoretical Physics}\\
{\em I-34100 Trieste, Italy}\\

\end{center}
\vskip 2.5cm
\begin{center}
{\bf Abstract}
\end{center}

Using Dirac's approach to constrained dynamics, the
Hamiltonian formulation of regular higher order Lagrangians is
developed. The
conventional description of such systems due to
Ostrogradsky is recovered.
However, unlike the latter, the present analysis
yields in a transparent manner the local
structure of the associated phase space and its
local sympletic geometry, and is
of direct application to {\em constrained\/}
higher order Lagrangian systems
which are beyond the scope of Ostrogradsky's approach.

\vskip 4cm

\end{titlepage}

\newpage

\section{Introduction}

Canonical and path integral
quantisations of systems whose dynamics is described
by higher order Lagrangians---namely by Lagrangians involving
time derivatives of the degrees of freedom
of order at least two---is an issue which by far has not been
developed to the same extent\footnote{The classical
analysis of such higher order regular or singular systems is available
to some extent in the recent literature\cite{Misc}.}
as for systems whose Lagrangian only
depends on the coordinates and their
velocities\cite{Govreview}. Nevertheless, there
do exist systems of physical interest described by such higher order
Lagrangians, the most popular examples being
perhaps higher order regularisations of quantum gauge
field theories and so-called rigid strings\cite{Polyakov,Klein}
or rigid particles\cite{Rigpart}.
In fact, these examples involve the additional
complication that they possess local symmetries, leading therefore
to constraints generating these gauge invariances on
phase space.

As is well known, there does exist a generalisation of the ordinary
Hamiltonian formulation in the case of higher order Lagrangians, which is
due to Ostrogradsky\cite{Ostro}. However, on the one hand,
Ostrogradsky's approach is implicitly
restricted to {\em non constrained\/} systems---which in
particular {\em do not\/} possess local gauge invariances---,
thus rendering this approach inapplicable to most, if not all physical
systems of present fundamental interest. On the other hand,
in Ostrogradsky's construction the structure of phase space and
in particular of its local symplectic geometry is not immediately
transparent, an obvious source of possible confusion when considering
canonical or path integral quantisations of such systems.

This note discusses how both problems can be resolved
within the
well established context of constrained systems\cite{Govreview} described by
Lagrangians depending on coordinates and velocities only.
Well known and powerful techniques become then immediately
available, rendering the necessity of a separate discussion of
the quantisation of higher order systems---including
constrained ones, and thus in particular their BRST quantisation---void
of any justification. {\em Any\/} higher order system
can {\em always\/} be cast in the form of an ordinary constrained
system, namely one whose Lagrangian is a function
only of first order time derivatives of the degrees of freedom, but
not of time derivatives of higher order.

That such a reduction of higher order Lagrangians is possible was
indicated already previously\cite{Pons,Gov2}.
As should be clear, it suffices for
this purpose to introduce auxiliary degrees of freedom associated
to each of the successive time derivatives of the original coordinates
of the system. In effect,
the canonical quantisation of rigid particles
has already used\cite{Rigpart} the same idea, thus in a situation where
Ostrogradsky's approach is not applicable as such.

The present note is organised as follows. In the next section, Ostrogradsky's
construction is briefly considered. Sect.3 describes how
any higher order Lagrangian system can be cast into the form
of a constrained system whose Lagrangian involves only first order
time derivatives of the degrees of freedom. The canonical Hamiltonian
description of the auxiliary system is then addressed in Sect.4 while
its equivalence with Ostrogradsky's formulation is established in
Sect.5. Further comments are presented in the Conclusion.

\section{Ostrogradsky's Construction}

Let us consider a system with degrees of freedom $x_n(t)$
$(n=1,2,\cdots)$, $t$ being the time evolution parameter
of the system. Although the present analysis assumes that these
coordinates are commuting variables, and that the index
$n$ takes a finite
or an infinite number of discrete values, it should be clear that
exactly the same considerations and the same conclusions
as those developed hereafter
are applicable to commuting and
anticommuting degrees of freedom, as well as to an infinite
non countable set of coordinates. The former case is
that of bosonic and fermionic
types of degrees of freedom, and the latter
typically that of field
theories. All the conclusions established in the present note
are thus valid in {\em complete generality\/},
for {\em any\/} system described
by some  higher order Lagrangian. The restriction to a {\em discrete\/}
set of {\em commuting\/}
degrees of freedom is only one of ease of presentation.
Moreover, in a first reading of the paper it might be useful to consider
the case of only one degree of freedom\cite{Pons}, namely ignore
the index $n$ altogether.

These remarks having been made, let us assume that the dynamical
time evolution of the system is determined from the variational
principle being applied to the action functional associated to some
time independent Lagrange function
\begin{equation}
L_0\left(x_n,\dot{x}_n,\ddot{x}_n,\cdots,x_n^{(m_n)}\right)\ .
\end{equation}
Here, $(m_n\geq 1)$ $(n=1,2,\cdots)$ is the maximal order of all time
derivatives of the coordinate $x_n$ $(n=1,2,\cdots)$
appearing in the Lagrangian. In particular, the discussion of
this paper includes the familiar case when $(m_n=1)$ for all
degrees of freedom. Throughout the analysis, it might be interesting
to consider the special case $(m_n=1)$ $(n=1,2,\cdots)$ to see how well-known
results are recovered from the present general situation.

Note that the Lagrange function is assumed to depend on at least the
first order time derivative of {\em each\/} degree of freedom
$x_n$. Otherwise, one would have to deal with some of the equations
of motion being actually constraints, a situation not considered
by Ostrogradsky. Moreover, the restriction to time {\em independent\/}
Lagrange functions is again for reasons of convenience rather than of
principle. Time dependent higher order Lagrangians can also be analysed
along the same lines as developed hereafter.

Considering the variational principle, it is clear that the Euler-Lagrange
equations of motion of the system are given by
\begin{equation}
\sum_{k_n=0}^{m_n}(-1)^{k_n}
\left(\frac{d}{dt}\right)^{k_n}
\frac{\partial L_0}{\partial x_n^{(k_n)}}=0\ ,\ \ n=1,2,\cdots\ .
\label{eq:EL0}
\end{equation}
Following Ostrogradsky's lead\cite{Ostro} and in order to simplify the
expression of these equations, let us introduce quantities
$p_{n,\alpha_n}$ $(\alpha_n=0,1,\cdots,m_n-1)$ defined recursively by
\begin{equation}
p_{n,i_n-1}=\frac{\partial L_0}{\partial x_n^{(i_n)}}
-\frac{d}{dt}p_{n,i_n}\ ,\ \ i_n=1,2,\cdots,m_n-1\ ,
\label{eq:recur1}
\end{equation}
with the initial value
\begin{equation}
p_{n,m_n-1}=\frac{\partial L_0}{\partial x_n^{(m_n)}}\ .
\label{eq:recur2}
\end{equation}
The Euler-Lagrange equations of motion in (\ref{eq:EL0}) take then
the simpler compact form
\begin{equation}
\frac{\partial L_0}{\partial x_n}-\frac{d}{dt}p_{n,0}=0\ ,
\ \ n=1,2,\cdots\ ,
\end{equation}
which are very suggestive of Hamiltonian equations of motion.
Note how these expressions generalise the familiar standard definitions
in the case when $(m_n=1)$ for all degrees of freedom
$x_n$ $(n=1,2,\cdots)$.

In order to reveal a possible Hamitonian description in the general
case when the integers $m_n$ take arbitrary finite values,
it is useful to consider the differential of the Lagrange function $L_0$,
in which the
definitions of the variables $p_{n,\alpha_n}$ are included.
A little calculation then leads to the identity,
\begin{displaymath}
d\left(\sum_n\sum_{\alpha_n=0}^{m_n-1}
x_n^{(\alpha_n+1)}p_{n,\alpha_n}-L_0\right)={\hspace{190pt}}
\end{displaymath}
\begin{equation}
=\sum_n\sum_{\alpha_n=0}^{m_n-1}\left[
x_n^{(\alpha_n+1)}dp_{n,\alpha_n}
-dx_n^{(\alpha_n)}\dot{p}_{n,\alpha_n}\right]
-\sum_n dx_n\left[\frac{\partial L_0}{\partial x_n}
-\dot{p}_{n,0}\right]\ .
\label{eq:identity}
\end{equation}
Note that the last sum in the r.h.s.
of this expression is a combination of the
Euler-Lagrange equations of motion of the system.

The meaning of this result is as follows. Consider the quantity
defined by
\begin{equation}
H=\sum_n\sum_{\alpha_n=0}^{m_n-1}\dot{x}_n^{(\alpha_n)}p_{n,\alpha_n}-L_0\ ,
\label{eq:H0}
\end{equation}
with the variables $p_{n,\alpha_n}$ determined by the recursion
relations in (\ref{eq:recur1}) and (\ref{eq:recur2}).
Since these variables involve the coordinates
$x_n$ and their time derivatives up to a certain order which is different
for each of the coordinates $x_n$ and each of the variables $p_{n,\alpha_n}$,
so would {\em a priori\/} the quantity $H$ given in (\ref{eq:H0}).
However, the identity (\ref{eq:identity}) establishes that this
dependence is in fact rather specific, namely only through a
dependence of the variables $x_n^{(\alpha_n)}$ and
$p_{n,\alpha_n}$ $(\alpha_n=0,1,\cdots,m_n-1)$ themselves,
\begin{equation}
H\left(x_n^{(\alpha_n)},p_{n,\alpha_n}\right)\ ,
\ \ \alpha_n = 0,1,\cdots,m_n-1\ .
\end{equation}
Note that this conclusion is valid irrespective of whether the
Lagrangian $L_0\left(x_n,\dot{x}_n,\cdots,x_n^{(m_n)}\right)$ leads
to constraints or not.

The identity (\ref{eq:identity}) also shows that
the Euler-Lagrange equations of motion (\ref{eq:EL0}) are
equivalent to the set of equations
\begin{equation}
\dot{x}_n^{(\alpha_n)}=\frac{\partial H}{\partial p_{n,\alpha_n}}\ , \ \
\dot{p}_{n,\alpha_n}=-\frac{\partial H}{\partial x_n^{(\alpha_n)}}\ ,
\ \ \alpha_n=0,1,\cdots,m_n-1\ .
\label{eq:HEM0}
\end{equation}
In other words, the system of higher order Lagrangian $L_0$
has been cast in Hamiltonian form, with the variables
$\left(x_n^{(\alpha_n)}, p_{n,m_n}\right)$ being canonically conjugate
pairs. Let us thus recapitulate.

Given the Lagrange function $L_0\left(x_n,\dot{x}_n,\cdots,x_n^{(m_n)}\right)$,
first one introduces the conjugate momenta $p_{n,m_n-1}$ defined by
\begin{equation}
p_{n,m_n-1}=\frac{\partial L_0}{\partial x_n^{(m_n)}}
\left(x_n,\dot{x}_n,\cdots,x_n^{(m_n)}\right)\ .
\label{eq:conj1}
\end{equation}
Regarding the variables $\left(x_n,\dot{x}_n,\cdots,x_n^{(m_n-1)}\right)$
as independent, the relations (\ref{eq:conj1}) may be inverted to give
\begin{equation}
x_n^{(m_n)}=\dot{x}_n^{(m_n-1)}=\dot{x}_n^{(m_n-1)}
\left(x_n,\dot{x}_n,\cdots,x_n^{(m_n-1)},p_{n,m_n}\right)\ .
\label{eq:velocity}
\end{equation}
The remaining conjugate momenta $p_{n,i_n-1}$ $(i_n=1,2,\cdots,m_n-1)$ are
then determined by the recursion relations in (\ref{eq:recur1}).
However, these relations are {\em not\/} used in order to express
the variables $\dot{x}_n^{(i_n-1)}$ $(i_n=1,2,\cdots,m_n-1)$
in terms of $(x_n,\dot{x}_n,\cdots,x_n^{(i_n-1)})$, the
conjugate momenta $(p_{n,i_n},\cdots,p_{n,m_n-1})$
and time derivatives of the latter. Indeed in Ostrogradsky's
construction, the variables
$(x_n,\dot{x}_n,\cdots,x_n^{(m_n-1)})$ have to be considered
as being {\em independent\/}.
We shall come back
to this point shortly.

Once the expressions for the quantities $\dot{x}_n^{(m_n-1)}$
determined as in (\ref{eq:velocity}), the canonical Hamiltonian
of the system is defined by (\ref{eq:H0}), or equivalently,
\begin{displaymath}
H(x_n^{(\alpha_n)},p_{n,\alpha_n})={\hspace{280pt}}
\end{displaymath}
\begin{equation}
=\dot{x}_n^{(m_n-1)}p_{n,m_n-1}-
L_0\left(x_n,\dot{x}_n,\cdots,\dot{x}_n^{(m_n-1)}\right)
+\sum_n\sum_{i_n=1}^{m_n-1}x_n^{(i_n)}p_{n,i_n-1}\ ,
\label{eq:H0prime}
\end{equation}
in which only the substitutions for the variables $\dot{x}_n^{(m_n-1)}$
are performed, while the degrees of freedom $x_n^{(\alpha_n)}$ and
$p_{n,\alpha_n}$ $(\alpha_n=0,1,\cdots,m_n-1)$ are considered
as being independent.

Finally, the Hamiltonian equations of motion are given by (\ref{eq:HEM0}).
Introducing the fundamental Poisson brackets,
\begin{equation}
\left\{x_n^{(\alpha_n)},x_m^{(\alpha_m)}\right\}=0\ ,\ \
\left\{x_n^{(\alpha_n)},p_{m,\alpha_m}\right\}=\delta_{nm}
\delta_{\alpha_n \alpha_m}\ ,\ \
\left\{p_{n,\alpha_n},p_{m,\alpha_m}\right\}=0\ ,
\label{eq:Ostrobrackets}
\end{equation}
with $(\alpha_n=0,1,\cdots,m_n-1)$ and $(\alpha_m=0,1,\cdots,m_m-1)$
$(n,m=1,2,\cdots)$,
the equations (\ref{eq:HEM0}) take the Hamiltonian form
\begin{equation}
\dot{x}_n^{(\alpha_n)}=\left\{x_n^{(\alpha_n)},H\right\}\ ,\ \
\dot{p}_{n,\alpha_n}=\left\{p_{n,\alpha_n},H\right\}\ ,
\ \ \alpha_n=0,1,\cdots,m_n-1\ .
\label{eq:HEM0prime}
\end{equation}
In other words, the variables $\left(x_n^{(\alpha_n)},p_{n,\alpha_n}\right)$
$(\alpha_n=0,1,\cdots,m_n-1)$ are pairs of conjugate degrees
of freedom, thus defining the phase space of the system and its local
symplectic structure.

Obviously, certain comments are in order. It is clear that
Ostrogradsky's construction is applicable only to those
higher order Lagrangians for which the inversions required in the
determination of the quantities $\dot{x}_n^{(m_n-1)}$
are non degenerate. Namely,
the Lagrangian $L_0$ {\em cannot\/} lead to constraints of any kind.
{\em Constrained\/} higher order Lagrangians are beyond
the scope of Ostrogradsky's approach.

Another issue with the present construction is the risk of confusion
which arises when dealing with the variables $x_n^{(\alpha_n)}$ and their
first order time derivatives $\dot{x}_n^{(\alpha_n)}$
$(\alpha_n=0,1,\cdots,m_n-1)$, a situation which becomes
even the more acute
when considering canonical or path integral quantisations
of such systems. As emphasized above, only the first order time derivatives
of the variables $x_n^{(m_n-1)}$ are to be solved for in terms of the
conjugate momenta $p_{n,m_n-1}$ and the variables
$x_n^{(\alpha_n)}$ $(\alpha_n=0,1,\cdots,m_n-1)$, the latter
considered to be independent of one another rather than being simply
time derivatives of order $\alpha_n$ of the coordinates $x_n$.
It is in this manner only that the canonical Hamiltonian defined
in (\ref{eq:H0prime}) can be made a function of the pairs
of conjugate degrees of freedom $(x_n^{(\alpha_n)},p_{n,\alpha_n})$.
To illustrate the possible confusion which might arise when this point is
not fully appreciated, the reader is invited to consider a simple
example in the case of a single degree of freedom $x(t)$, such as,
\begin{equation}
L_0(x,\dot{x},\ddot{x})=\frac{1}{2}ax\ddot{x}^2-\frac{1}{2}bx\dot{x}^2\ ,
\label{eq:example}
\end{equation}
with $a$ and $b$ being arbitrary constant parameters.
If one attempts solving
both for $\ddot{x}$ and for $\dot{x}$ in terms of $x$ and $p_0$ and
$p_1$, there appear in the canonical Hamiltonian time derivative terms
of the conjugate momentum $p_1$! It is thus important to develop
Ostrogradsky's construction precisely in the manner emphasized above,
keeping the variables $x$ and $\dot{x}$ as independent, and
inverting only for $\ddot{x}$ in terms of $x$, $\dot{x}$ and $p_1$.

Nevertheless, when solving the Hamiltonian equations of motion
(\ref{eq:HEM0prime}) for the degrees of freedom $x_n(t)$, it becomes
necessary, {\em after having computed the Poisson brackets\/},
to impose the condition that the variables $x_n^{(i_n)}$ are
time derivatives of order $i_n$ $(i_n=1,2,\cdots,m_n-1)$
of the coordinates $x_n(t)$.

It is clear that both issues are solved at once by emphasizing
{\em explicitly\/}
the fact that in the Hamiltonian approach---hence also when
considering canonical and path integral quantisations of
such systems---, all variables
$x_n^{(\alpha_n)}$ are to be considered as being {\em independent\/}.
This is readily achieved by introducing {\em independent auxiliary\/}
degrees of freedom, each corresponding to a time derivative
of a given order of one of the original degrees of freedom.
The {\em same\/} system
can then be described in terms of an extended Lagrangian including
a dependence on the auxiliary degrees of freedom, such that
time derivatives of first order only are involved. In this manner, one
is brought back\cite{Pons,Gov2} into the realm of the usual type of dynamical
systems for which most powerful techniques are available, with the
additional advantage that {\em constrained\/} higher order
Lagrangians do not need to be considered on a separate basis
any longer.

\section{The Auxiliary Lagrangian}

Given a system of degrees of freedom $x_n(t)$ $(n=1,2,\cdots)$
with Lagrange function $L_0(x_n,\dot{x}_n,\cdots,x_n^{(m_n)})$
$(m_n\geq 1)$---{\em be it regular or not\/}---, let us introduce new
{\em independent\/} variables $q_{n,\alpha_n}(t)$
$(\alpha_n=0,1,\cdots,m_n-1)$ such that the following recursion
relations would hold,
\begin{equation}
q_{n,i_n}=\dot{q}_{n,i_n-1}\ ,\ \ i_n=1,2,\cdots,m_n-1\ ,
\label{eq:recq1}
\end{equation}
with the initial value
\begin{equation}
q_{n,0}=x_n\ .
\label{eq:recq2}
\end{equation}
Clearly, the variables $q_{n,i_n}$ $(i_n=1,2,\cdots,m_n-1)$ would
then correspond to the
time derivatives $x_n^{(i_n)}$ of order $i_n$
of the coordinates $x_n$, the latter
being identical to the coordinates $q_{n,0}$.

In order to inforce
the relations (\ref{eq:recq1}) and (\ref{eq:recq2}) for
the {\em independent\/} variables $q_{n,\alpha_n}$, additional
Lagrange multipliers $\mu_{n,i_n}(t)$ $(i_n=1,2,\cdots,m_n-1)$
are introduced. The variables $(q_{n,\alpha_n},\mu_{n,i_n})$
thus determine the set of {\em independent\/} degrees of freedom of
the extended Lagrangian system,
with $(q_{n,i_n},\mu_{n,i_n})$ $(i_n=1,2,\cdots,m_n-1)$ being
auxiliary degrees of freedom as compared to the original coordinates
$\left(x_n(t)=q_{n,0}(t)\right)$.
The auxiliary Lagrange function of this extended description
of the system is given by
\begin{displaymath}
L(q_{n,\alpha_n},\dot{q}_{n,\alpha_n},\mu_{n,i_n})={\hspace{270pt}}
\end{displaymath}
\begin{equation}
=L_0(q_{n,0},q_{n,1},\cdots,q_{n,m_n-1},
\dot{q}_{n,m_n-1})+\sum_n\sum_{i_n=1}^{m_n-1}
\left(q_{n,i_n}-\dot{q}_{n,i_n-1}\right)\mu_{n,i_n}\ .
\label{eq:L1}
\end{equation}
Note that as advertised, the auxiliary Lagrangian $L$ involves
only {\em first\/} order time derivatives of the extended set
of degrees of freedom. Obviously, due to the presence of the
Lagrange multipliers $\mu_{n,i_n}$, the Lagrange function
$L$ in (\ref{eq:L1}) defines a constrained system, to which
the usual analysis\cite{Dirac,Govreview} of constrained dynamics is applicable.

Before turning to that important issue however, let us first establish the
equivalence of the auxiliary Lagrangian with the original
formulation of the system determined by the Lagrangian $L_0$.
Applied to $L$ in (\ref{eq:L1}), the variational principle
leads to the following equations of motion for the Lagrange
multipliers $\mu_{n,i_n}$,
\begin{equation}q_{n,i_n}=\dot{q}_{n,i_n-1}\ ,\ \ i_n=1,2,\cdots,m_n-1\ ,
\label{eq:EMmu}
\end{equation}
while for the degrees of freedom $q_{n,i_n}$ $(i_n=1,2,\cdots,m_n-1)$,
one obtains,
\begin{equation}
\mu_{n,j_n}=-\frac{\partial L_0}{\partial q_{n,j_n}}
-\frac{d}{dt}\mu_{n,j_n+1}\ ,\ \ j_n=1,2,\cdots,m_n-2\ ,
\label{eq:EMqni1}
\end{equation}
and
\begin{equation}
\mu_{n,m_n-1}=-\frac{\partial L_0}{\partial q_{n,m_n-1}}
+\frac{d}{dt}\frac{\partial L_0}{\partial\dot{q}_{n,m_n-1}}\ .
\label{eq:EMqni2}
\end{equation}
Finally, the equations of motion for $q_{n,0}$ are
\begin{equation}
\frac{\partial L_0}{\partial q_{n,0}}+\dot{\mu}_{n,1}=0\ .
\label{eq:EMqn0}
\end{equation}

Note that the latter equations are in fact the actual equations
of motion of the system. Indeed, all the other equations for
$q_{n,i_n}$ and $\mu_{n,i_n}$ are constraint equations which determine
the auxiliary degrees of freedom $q_{n,i_n}$ in terms of
successive time derivatives of the original coordinates
$(q_{n,0}=x_n)$, as well as
the Lagrange multipliers $\mu_{n,i_n}$ in terms of successive
partial derivatives of the Lagrange function $L_0$.
By substitution in (\ref{eq:EMqn0})
of the successive definitions of the Lagrange
multipliers $\mu_{n,i_n}$, the equations
of motion for $q_{n,0}$ reduce to
\begin{equation}
\sum_{\alpha_n=0}^{m_n-1}(-1)^{\alpha_n}
\left(\frac{d}{dt}\right)^{\alpha_n}
\frac{\partial L_0}{\partial q_{n,\alpha_n}}+
(-1)^{m_n}\left(\frac{d}{dt}\right)^{m_n}
\frac{\partial L_0}{\partial \dot{q}_{n,m_n-1}}=0\ .
\end{equation}
Upon the substitution of the recursion relations (\ref{eq:EMmu}),
one then indeed recovers the original Euler-Lagrange equations
of motion in (\ref{eq:EL0}).
Hence, the complete
equivalence between the auxiliary formulation of the system
and the original one based on the higher order Lagrange
function $L_0\left(x_n,\dot{x}_n,\cdots,x_n^{(m_n)}\right)$ is established.

\section{The Hamiltonian Formulation}

Given the auxiliary Lagrangian formulation of higher order systems
of the previous section,
let us apply to it the ordinary analysis\cite{Govreview}
of constraints in order
to develop its Hamiltonian description.
The momenta canonically
conjugate to the degrees of freedom $q_{n,\alpha_n}$
$(\alpha_n=0,1,\cdots,m_n-1)$ and
$\mu_{n,i_n}$ $(i_n=1,,2\cdots,m_n-1)$
are of course defined by, respectively,
\begin{equation}
p_{n,\alpha_n}=\frac{\partial L}{\partial \dot{q}_{n,\alpha_n}}\ ,\ \
\pi_{n,i_n}=\frac{\partial L}{\partial\dot{\mu}_{n,i_n}}\ .
\end{equation}

However, the phase space degrees of freedom
$(q_{n,\alpha_n},p_{n,\alpha_n};\mu_{n,i_n},\pi_{n,i_n})$ are
not all independent. In fact, the system possesses the following
primary constraints,
\begin{equation}
\Phi_{n,i_n}=0\ ,
\ \ \pi_{n,i_n}=0\ ,\ \
i_n=1,2,\cdots,m_n-1\ ,
\label{eq:Constraints}
\end{equation}
where
\begin{equation}
\Phi_{n,i_n}\equiv p_{n,i_n-1}+\mu_{n,i_n}\ , \ \
i_n=1,2,\cdots,m_n-1\ .
\end{equation}
Both sets of primary constraints follow from the particular way
in which the auxiliary degrees of freedom are introduced in the
definition of the extended
Lagrange function $L$ in (\ref{eq:L1}).
The primary constraints obey the algebra of Poisson brackets
\begin{equation}
\left\{\Phi_{n,i_n},\Phi_{m,i_m}\right\}=0\ ,\ \
\left\{\Phi_{n,i_n},\pi_{m,i_m}\right\}=
\delta_{nm}\delta_{i_ni_m}\ ,\ \
\left\{\pi_{n,i_n},\pi_{m,i_m}\right\}=0\ ,
\label{eq:algebraconstraints}
\end{equation}
with $(i_n=1,2,\cdots,m_n-1)$ and $(i_m=1,2,\cdots,m_m-1)$
$(n,m=1,2,\cdots)$,
showing therefore already at this stage that the primary constraints are
certainly also second class constraints.

Among all conjugate momenta, $p_{n,m_n-1}$ certainly
play a distinguished role since on the one hand, they are the only
ones not involved in any of the primary constraints above, and on the other
hand, their conjugate coordinates $q_{n,m_n-1}$ are the only variables
whose first order time derivatives do appear in the original
Lagrange function $L_0$. Indeed, we have
\begin{equation}
p_{n,m_n-1}=\frac{\partial L_0}{\partial\dot{q}_{n,m_n-1}}
\left(q_{n,0},q_{n,1},\cdots,q_{n,m_n-1},\dot{q}_{n,m_n-1}\right)\ .
\label{eq:pnmn-1}
\end{equation}
As in Ostrogradsky's approach, let us then assume that for {\em fixed\/}
values of $q_{n,\alpha_n}$ $(\alpha=0,1,\cdots,m_n-1)$, these relations
are invertible, leading therefore to the velocities,
\begin{equation}
\dot{q}_{n,m_n-1}=\dot{q}_{n,m_n-1}
\left(q_{n,\alpha_n},p_{n,m_n-1}\right)\ .
\end{equation}
In other words, given fixed values for
$q_{n,i_n-1}$ $(i_n=1,2,\cdots,m_n-1)$,
the dynamical system of degrees of freedom $q_{n,m_n-1}$ with
Lagrange function
$L_0\left(q_{n,0},q_{n,1},\cdots,q_{n,m_n-1},\dot{q}_{n,m_n-1}\right)$
is assumed to be a {\em regular\/} system, namely {\em not leading\/}
to any constraints for the conjugate momenta $p_{n,m_n-1}$.
Consequently, the constraints in (\ref{eq:Constraints}) determine
the full set of primary contraints in the extended formalism of the
higher order Lagrangian $L_0\left(x_n,\dot{x}_n,\cdots,x_n^{(m_n)}\right)$.

The distinguished role of the conjugate variables
$(q_{n,m_n-1},p_{n,m_n-1})$ justifies the definition of the
{\em restricted Legendre transform\/} of
$L_0\left(q_{n,\alpha_n},\dot{q}_{n,m_n-1}\right)$,
leading to the {\em restricted canonical Hamiltonian\/},
\begin{equation}
\overline{H}_0(q_{n,\alpha_n},p_{n,m_n-1})=
\sum_n\dot{q}_{n,m_n-1}p_{n,m_n-1}-
L_0\left(q_{n,\alpha_n},\dot{q}_{n,m_n-1}\right)\ .
\label{eq:H0restricted}
\end{equation}
In the same way as was established for the Hamiltonian $H$ in (\ref{eq:H0}),
note that the restricted Hamiltonian $\overline{H}_0$ is a function of the
variables $(q_{n,\alpha_n},p_{n,m_n-1})$ only,
{\em irrespective of whether the relations
(\ref{eq:pnmn-1}) are invertible
or not\/}, namely irrespective of whether
$L_0\left(q_{n,0},q_{n,1},\cdots,q_{n,m_n-1},\dot{q}_{n,m_n-1}\right)$
defines a regular system in the coordinates $q_{n,m_n-1}$ or
not\cite{Gov}. In the present discussion, the assumption of regularity
is necessary only in order that no further primary constraints
beyond those in (\ref{eq:Constraints}) appear in the analysis.

In terms of the definitions and the primary constraints above, the
ca\-no\-ni\-cal Hamiltonian of the extended system,
\begin{equation}
H_0=\sum_n\sum_{\alpha_n}^{m_n-1}\dot{q}_{n,\alpha_n}p_{n,\alpha_n}
+\sum_n\sum_{i_n=1}^{m_n-1}\dot{\mu}_{n,i_n}\pi_{n,i_n}-L\ ,
\end{equation}
is readily found to be given by
\begin{equation}
H_0\left(q_{n,\alpha_n},p_{n,\alpha_n};\mu_{n,i_n}\right)=
\overline{H}_0\left(q_{n,\alpha_n},p_{n,m_n-1}\right)
-\sum_n\sum_{i_n=1}^{m_n-1}\mu_{n,i_n}q_{n,i_n}\ .
\label{eq:concanH0}
\end{equation}
However, as is well known\cite{Govreview}, due to the presence of constraints,
the Hamiltonian generating the genuine time evolution of the system
under which the constraints are preserved, is in general given by the
canonical Hamiltonian $H_0$ and a linear combination of the constraints.
Thus in the present case, the would-be Hamiltonian is of the form,
\begin{equation}
H_*=H_0+\sum_n\sum_{i_n=1}^{m_n-1}\left[
\lambda_{n,i_n}^{(1)}\Phi_{n,i_n}+
\lambda_{n,i_n}^{(2)}\pi_{n,i_n}\right]\ ,
\end{equation}
with $\lambda_{n,i_n}^{(1)}$ and $\lambda_{n,i_n}^{(2)}$ being
Lagrange multipliers for the constraints.
Consistent time evolution of the primary constraints $\Phi_{n,i_n}$
and $\pi_{n,i_n}$ then imposes the relations
\begin{equation}
\lambda_{n,i_n}^{(1)}=q_{n,i_n}\ ,\ \ i_n=1,2,\cdots,m_n-1\ ,
\end{equation}
as well as
\begin{equation}
\lambda_{n,1}^{(2)}=\frac{\partial\overline{H}_0}{\partial q_{n,0}}\ ,\ \
\lambda_{n,j_n}^{(2)}=\frac{\partial\overline{H}_0}{\partial q_{n,j_n-1}}
-\mu_{n,j_n-1}\ ,\ \ j_n=2,3,\cdots,m_n-1\ .
\end{equation}
Consequently, the extended formulation of the system does not possess
se\-con\-dary constraints, while its extended Hamiltonian reduces to
\begin{displaymath}
H_*=\overline{H}_0+\sum_n\sum_{i_n=1}^{m_n-1}
q_{n,i_n}p_{n,i_n-1}+{\hspace{100pt}}
\end{displaymath}
\begin{equation}
+\sum_n\pi_{n,1}\frac{\partial\overline{H}_0}{\partial q_{n,0}}+
\sum_n\sum_{j_n=2}^{m_n-1}\pi_{n,j_n}
\left[\frac{\partial\overline{H}_0}{\partial q_{n,j_n-1}}
-\mu_{n,j_n-1}\right]\ .
\end{equation}

However, as already pointed out previously,
all primary constraints $\Phi_{n,i_n}$ and
$\pi_{n,i_n}$ are second class, and may thus be solved for explicitly
provided the canonical Poisson brackets are traded for appropriate
Dirac brackets\cite{Dirac,Govreview}.
Choosing to solve the constraints in terms of
\begin{equation}
\mu_{n,i_n}=-p_{n,i_n-1}\ ,\ \ \pi_{n,i_n}=0\ ,
\ \ i_n=1,2,\cdots,m_n-1\ ,
\end{equation}
the reduced phase space degrees of freedom are then simply
$(q_{n,\alpha_n},p_{n,\alpha_n})$ $(\alpha_n=0,1,\cdots,m_n-1)$.
On the other hand, given the algebra (\ref{eq:algebraconstraints}) of
constraints, the Dirac brackets of the reduced Hamiltonian description
are easily seen to remain canonical
$(\alpha_n=0,1,\cdots,m_n-1;\ \alpha_m=0,1,\cdots,m_m-1)$,
\begin{equation}
\left\{q_{n,\alpha_n},q_{m,\alpha_m}\right\}_D=0\ ,\ \
\left\{q_{n,\alpha_n},p_{m,\alpha_m}\right\}_D=
\delta_{nm}\delta_{\alpha_n\alpha_m}\ ,\ \
\left\{p_{n,\alpha_n},p_{m,\alpha_m}\right\}_D=0\ .
\label{eq:Diracbrackets}
\end{equation}

The constrained description of higher order Lagrangian systems has thus lead to
the following Hamiltonian formulation. The local phase space coordinates
are the canonically conjugate pairs of degrees of freedom
$(q_{n,\alpha_n},p_{n,\alpha_n})$ $(\alpha_n=0,1,\cdots,m_n-1)$, with
the local symplectic structure determined by the Dirac brackets
in (\ref{eq:Diracbrackets}). Time evolution in phase space is specified
through these brackets by the extended Hamiltonian,
\begin{equation}
H_E\left(q_{n,\alpha_n},p_{n,\alpha_n}\right)=
\overline{H}_0\left(q_{n,\alpha_n},p_{n,m_n-1}\right)+
\sum_n\sum_{i_n=1}^{m_n-1}q_{n,i_n}p_{n,i_n-1}\ .
\label{eq:extendedHamiltonian}
\end{equation}
In particular, the fundamental equations of motion are
$(i_n=1,2,\cdots,m_n-1)$,
\begin{equation}
\dot{q}_{n,i_n-1}=q_{n,i_n}\ ,\ \
\dot{q}_{n,m_n-1}=\frac{\partial\overline{H}_0}{\partial p_{n,m_n-1}}\ ,
\end{equation}
as well as
\begin{equation}
\dot{p}_{n,0}=-\frac{\partial\overline{H}_0}{\partial q_{n,0}}\ ,\ \
\dot{p}_{n,i_n}=-\frac{\partial\overline{H}_0}{\partial q_{n,i_n}}
-p_{n,i_n-1}\ .
\label{eq:EMpnalphan}
\end{equation}
However, the restricted canonical Hamiltonian $\overline{H}_0$ is
such that
\begin{equation}
\frac{\partial\overline{H}_0}{\partial q_{n,\alpha_n}}
\left(q_{n,\alpha_n},p_{n,m_n-1}\right)=
-\frac{\partial L_0}{\partial q_{n,\alpha_n}}
\left(q_{n,\alpha_n},
\dot{q}_{n,m_n-1}(q_{n,\alpha_n},p_{n,m_n-1})\right)\ ,
\end{equation}
so that the equations (\ref{eq:EMpnalphan}) for $p_{n,\alpha_n}$ are
equivalent to
\begin{equation}
\frac{\partial L_0}{\partial q_{n,0}}-\frac{d}{dt}p_{n,0}=0\ ,
\label{eq:HEMpn0}
\end{equation}
with
\begin{equation}
p_{n,i_n-1}=\frac{\partial L_0}{\partial q_{n,i_n}}-\frac{d}{dt}p_{n,i_n}\ ,
\ \ i_n=1,2,\cdots,m_n-1\ .
\label{eq:HEMpni}
\end{equation}

Expressed in this manner, it is clear how these Hamiltonian equations
of motion are indeed equivalent to the Euler-Lagrange equations
(\ref{eq:EMmu}) to (\ref{eq:EMqn0}) of Sect.3, when the constraints
$(\mu_{n,i_n}=-p_{n,i_n-1})$ are accounted for. Indeed,
(\ref{eq:HEMpn0}) determine the actual equations of motion of the
system and are equivalent to (\ref{eq:EMqn0}). The conjugate momenta
$p_{n,0}$ are defined recursively through
the equations (\ref{eq:HEMpni}), which
are equivalent to (\ref{eq:EMqni1}) and (\ref{eq:EMqni2}), given
the momenta $p_{n,m_n-1}$. Finally, the latter quantities, which are
absent of course in the Lagrangian equations of motion except through
their definition as
\begin{equation}
p_{n,m_n-1}=\frac{\partial L_0}{\partial\dot{q}_{n,m_n-1}}
\left(q_{n,\alpha_n},\dot{q}_{n,m_n-1}\right)\ ,
\label{eq:conjmom}
\end{equation}
are determined implicitly by the Hamiltonian equations of motion,
\begin{equation}
\dot{q}_{n,m_n-1}=\frac{\partial\overline{H}_0}{\partial p_{n,m_n-1}}
\left(q_{n,\alpha_n},p_{n,m_n-1}\right)\ .
\end{equation}
Since the Lagrangian $L_0\left(q_{n,\alpha_n},\dot{q}_{n,m_n-1}\right)$
is assumed to be regular in the coordinates
$q_{n,m_n-1}$, this latter relation
is indeed invertible, leading back to the relation (\ref{eq:conjmom}).
It is in this way that the present Hamiltonian equations of motion are
equivalent to the Euler-Lagrange equations under the Lagrangian
reduction, namely the reduction of conjugate momenta $p_{n,\alpha_n}$ in terms
of the coordinates $q_{n,\alpha_n}$ and their velocities
$\dot{q}_{n,\alpha_n}$. Therefore, since the auxiliary Lagrangian formulation
was shown to reproduce the Euler-Lagrange equations of the higher order
Lagrangian, the present Hamiltonian construction is established to
be equivalent to the original description of the system as well.

\section{Ostrogradsky's Approach Revisited}

The equivalence of the results obtained through the analysis of constraints
applied to the auxiliary formulation of {\em regular\/} higher order
Lagrangian systems with Ostrogradsky's construction is now obvious.

In the latter approach at the Hamiltonian level, the successive time
derivatives $x_n^{(\alpha_n)}$ $(\alpha_n=0,1,\cdots,m_n-1)$ of
the degrees of freedom $x_n$ {\em have to be considered as being
independent\/}. In the constrained formulation, these variables correspond
to the {\em independent\/} auxiliary coordinates $q_{n,\alpha_n}$,
with in particular $(q_{n,0}=x_n)$.
In addition, the fundamental brackets (\ref{eq:Ostrobrackets}) in
Ostrogradsky's formulation are identical to the canonical Dirac
brackets(\ref{eq:Diracbrackets}) of the reduced phase space
degrees of freedom $(q_{n,\alpha_n},p_{n,\alpha_n})$.

Finally, it is clear that the extended Hamiltonian $H_E$ in
(\ref{eq:extendedHamiltonian}) is {\em identical\/} to
Ostrogradsky's canonical Hamiltonian $H$ in (\ref{eq:H0prime}).
In particular, note how in the definition of the latter quantity,
the restricted canonical Hamiltonian $\overline{H}_0$ defined
in (\ref{eq:H0restricted}) appears naturally, indeed emphasizing once again
the distinguished role played by the time derivatives
$x_n^{(m_n)}$ of maximal order of the degrees of freedom $x_n(t)$.

Therefore, the analysis of the previous section, based on the
auxiliary formulation of regular higher order Lagrangian systems
and Dirac's analysis of constraints, has recovered precisely Ostrogradsky's
Hamiltonian description of such systems.

\section{Conclusion}

This note has established the equivalence of the
Hamiltonian formulation of {\em regular\/} higher
order Lagrangian systems due to Ostrogradsky\cite{Ostro},
with a constrained auxiliary description\cite{Pons,Gov2}
of such systems in which time derivatives of degrees
of freedom of at most first order only are involved.
The latter approach offers the following advantages, however.

In Ostrogradsky's construction, time derivatives
of the coordinates of different
order have to be considered as being {\em independent\/}.
Such a situation is a possible source of confusion, especially at the
quantum level when translating Poisson brackets into (anti)commutation
relations for the fundamental quantum operators. Indeed, it is not
always clear when to consider a time derivative of given order as
an independent variable or as the first order time derivative of
some other variable in Ostrogradsky's phase space. In the auxiliary
approach, this issue is avoided altogether {\em ab initio\/},
since {\em independent\/} auxiliary degrees of freedom are introduced
explicitly, each being associated with a time derivative of given
order of each of the original degrees of freedom. In this manner,
the local structure of phase space and its local symplectic geometry, is
made perhaps much more transparent than in Ostrogradsky's approach.

More importantly however, the auxiliary formulation presents the additional
advantage that the auxiliary Lagrangian depends on time derivatives
of first order only. Therefore, {\em any\/} higher order Lagrangian
system---{\em be it regular or not\/}---can always be brought into the
realm of those Lagrangian systems for which a wealth of
methods---classical and quantum---have been developed over the years.
Due to the presence of auxiliary degrees of freedom, the auxiliary formulation
always leads to constraints, requiring the techniques of constrained
dynamics\cite{Govreview}.

Finally, in contradistinction to Ostrogradsky's construction which
applies to {\em regular\/} higher order systems {\em only\/}, the
auxiliary formulation, being already a constrained one, does not require
to distinguish between regular and singular higher order Lagrangian
systems. Hence, the quantisation of such systems,
including the BRST quantisation of singular ones, {\em does not
necessitate a separate and generalised formalism not yet developed\/}.
All the readily available methods of ordinary constrained
quantisation---{\em and nothing more\/}---suffice for the Hamiltonian
formulation and the quantisation of {\em any\/} higher order
Lagrangian system. As this note has established, Ostrogradsky's
construction is thereby recovered exactly in the case of
regular systems. The case of singular systems however, is beyond the
scope of the latter approach, and the auxiliary formulation
then becomes unavoidable. In effect, precisely this method has been
applied already to rigid particles for example, with important
conclusions as to their quantum consistency\cite{Rigpart}.

\section*{Acknowledgements}

One of the authors (M.S.R) would like to thank
Professor Abdus Salam, the International Atomic agency,
UNESCO and the International
Centre for Theoretical Physics (Trieste) for support.

\end{document}